\documentclass[aps,rpl,preprint,groupedaddress]{revtex4}

\usepackage{graphicx}

\begin{document}

\title{Effects of quantum noises and noisy quantum operations on entanglement and special dense coding}
\author{Sylvanus Quek}
\affiliation{Hwa Chong Institution, 661 Bukit Timah Road, Singapore 269734, Singapore}

\author{Ziang Li}
\affiliation{Hwa Chong Institution, 661 Bukit Timah Road, Singapore 269734, Singapore}

\author{Ye Yeo}
\affiliation{Department of Physics, National University of Singapore, 10 Kent Ridge Crescent, Singapore 119260, Singapore}

\begin{abstract}
We show how, in the presence of quantum noises generated by non-commuting Lindblad operators, a maximally entangled Bell state may suffer entanglement 
sudden death (ESD).  Similarly, ESD may occur when a Bell state is under the action of a quantum operation generated by a Hamiltonian in the presence of 
a quantum noise, provided that the Hamiltonian does not commute with the corresponding Lindblad operator.  Practically this means noisy quantum 
operations could cause ESD during the encoding process in quantum communication protocols like special dense coding ($\cal S$).  Next, we show how in 
the presence of quantum noises which cause ESD, a Bell state may lose its capacity for $\cal S$ before ESD occurs.  Finally, we show that a Bell state 
may indeed fail to yield information transfer better than classically possible when the encoding operations of $\cal S$ are noisy, even though 
entanglement is not totally destroyed in the process.

\end{abstract}

\maketitle

\section{Introduction}
Quantum communication \cite{Nielsen} promises means of information transfer that are classically impossible.  For instance, superdense coding, first 
proposed by Bennett and Wiesner \cite{Bennett92}, allows Alice to send Bob two bits of classical information by the physical transfer of a single 
encoded qubit employing prior maximal entanglement associated with a Bell state $\chi_0$.  At the heart of such quantum superiority is quantum coherence 
of both the systems where information resides, and the operations that process information.  However, real systems and processors can never be perfectly 
isolated from the surrounding world.  Environmental influences always result in loss of coherence, or decoherence \cite{Breuer}.  In order to formulate 
effective strategies to address these issues in quantum communication, it is imperative to have a complete understanding of the effects of decoherence 
on protocols such as superdense coding.

A step in this direction was taken by Bose, {\it et al.} \cite{Bose} and Bowen \cite{Bowen}, who considered the case where the initial pairs of qubits 
being used for superdense coding are in a non-maximally entangled mixed state $\chi$.  In particular, Bowen derived the classical information capacity 
of superdense coding, through a noiseless channel, using arbitrary mixed states of two qubits.  If Alice and Bob share pairs of qubits in the state 
$\chi_{AB}$, and Alice is restricted to using unitary operators and sending her message as a product state of letters, then the classical information 
capacity of superdense coding is given by \cite{Bowen}
\begin{equation}\label{cap}
{\cal C}[\chi_{AB}] = 1 + S[\chi_B] - S[\chi_{AB}],
\end{equation}
where $\chi_B = {\rm tr}_A(\chi_{AB})$ and $S[\rho] = -{\rm tr}[\rho\log_2\rho]$ denotes the von Neumann entropy of the state $\rho$.  This is 
attainable using {\it special dense coding} ${\cal S}$, where Alice encodes a message by applying to her qubits unitary transformations from the set 
$\{\sigma_0, \sigma_1, \sigma_2, \sigma_3\}$ of the two-dimensional identity and three Pauli operators, with equal a priori probabilities of $1/4$.  The 
Pauli operators satisfy the commutation relations $[\sigma_m, \sigma_n] = 2i\epsilon_{lmn}\sigma_l$.

In Refs.\cite{Bose, Bowen}, presumably the non-maximally entangled mixed state $\chi$ is a result of transmission noise, described by ${\cal T}$.  For 
instance, Alice could have prepared pairs of qubits in a maximally entangled Bell state $\chi_0$ at time $t = 0$ and send one qubit from each pair to 
Bob via a noisy quantum channel.  If it takes time $t_0$ for the qubit to reach Bob, we have $\chi = {\cal T}(\chi_0, t_0)$.  However, it is supposed 
that Alice sends her encoded qubits to Bob via a noiseless channel.  Furthermore, it is assumed that Alice's encoding operations are perfect unitary 
ones.  The aim of these work is to study in what way the capacity to do superdense coding depend on the degree of entanglement of $\chi$.  To this end 
these assumptions may be justifiable, but they certainly do not hold in reality.  As a matter of fact, one of us has recently shown how noisy quantum 
operations could totally destroy entanglement in a finite time \cite{Yeo}.  This work is inspired by that of Yu and Eberly \cite{Yu}, where it has been 
shown that even under ideally weak phase and amplitude noise the entanglement associated with a generic class of entangled states goes abruptly to zero 
in a finite time and remains zero thereafter.  This phenomenon, called ``entanglement sudden death" (ESD) \cite{Yu}, has recently been confirmed 
experimentally \cite{Almeida, Eberly}.  It is shown, in Ref.\cite{Yeo}, that transmission noise together with noisy quantum operations may render an 
initially maximally entangled Bell state useless for standard teleportation \cite{Bennett93}, even though the channel state shared between Alice and Bob 
was entangled before Alice's Bell measurement.  This is thus obviously different from ESD.  In the light of these developments, it is worthwhile to 
analyze how other quantum communication protocols, like superdense coding, are affected by environmental effects during transmission, decoding, encoding, 
etc.  We focus on special dense coding ${\cal S}$ in this paper.

Our paper is organized as follows.  In Section II, we provide a detailed study of environmental effects on entanglement.  Firstly, we show how in the 
presence of quantum noises generated by non-commuting Lindblad operators, the entanglement associated with a maximally entangled Bell state may suffer 
sudden death.  Then, we show that ESD may also occur when a Bell state is under the action of a quantum operation generated by a Hamiltonian in the 
presence of a quantum noise, provided that the Hamiltonian does not commute with the corresponding Lindblad operator.  In Section III, a sketch of 
realistic special dense coding together with a brief review of Holevo function is given.  These set the stage for the presentation of our further 
results in Section IV.  Firstly, we show how when environmental influences during transmission alone are capable of causing ESD, they may result in a 
Bell state losing its capacity for $\cal S$ before ESD occurs.  Next, we show that Alice's noisy encoding operations alone could also yield an equally 
devastating impact on $\cal S$.  These results are in contrast to those for standard teleportation \cite{Yeo}, and clearly justify our investigation.  
We conclude in Section V.

\section{Noncommuting noises, Noisy quantum operations, and ESD}
Under the assumption that the environment is Markovian, the state $\rho$ of an open system evolves according to a quantum master equation \cite{Breuer}
\begin{equation}\label{master}
\frac{d}{dt}\rho = -i[H, \rho] + \sum_k\frac{1}{2}\gamma_k(2L_k\rho L^{\dagger}_k - L^{\dagger}_kL_k\rho - \rho L^{\dagger}_kL_k).
\end{equation}
Here, we set $\hbar = 1$.  $H$ is the system's Hamiltonian.  $\gamma_k$'s describe environmental coupling strengths.  $L_k$'s are the Lindblad operators.  
They are the generators of noise.  Alone, they purely cause states to lose their quantumness.  In the following, we consider a pair of qubits initially 
in the maximally entangled state $\chi_0 \equiv |\Psi^0_{\rm Bell}\rangle\langle\Psi^0_{\rm Bell}|$, where $|\Psi^0_{\rm Bell}\rangle = 
(|00\rangle + |11\rangle)/\sqrt{2}$.  We analyze the effects of noise and noisy operations on the entanglement of $\chi_0$.  For a two-qubit system $AB$, 
it is well-known that a necessary and sufficient condition for separability is that a matrix, obtained by partial transposition of its density operator, 
has non-negative eigenvalue(s) \cite{Peres}.  As a measure of the amount of entanglement associated with a given two-qubit state $\rho_{AB}$, we employ 
the negativity \cite{Vidal}
\begin{equation}
{\cal N}[\rho_{AB}] \equiv \max\left\{0, -2\sum_n\lambda_n\right\},
\end{equation}
where $\lambda_n$ is a negative eigenvalue of $\rho^{T_A}_{AB}$, the partial transposition of $\rho_{AB}$.

\subsection{Phase noise}
To set the stage, we begin with the situation where one of the particles, originally in the state $\chi_0$, is subject to a phase noise described by 
$L_{03} = \sigma_0 \otimes \sigma_3$ (or $L_{30} = \sigma_3 \otimes \sigma_0$).  Solving the master equation,
$$
\frac{d}{dt}\rho = \gamma_3[(\sigma_0 \otimes \sigma_3)\rho(\sigma_0 \otimes \sigma_3) - \rho].
$$
we find at time $t = t_0 > 0$:
\begin{equation}\label{diag0}
{\cal Z}(\chi_0, t_0) = 
\frac{1}{2}\left(\begin{array}{cccc}
1 & 0 & 0 & e^{-2\gamma_3 t_0} \\
0 & 0 & 0 & 0 \\
0 & 0 & 0 & 0 \\
e^{-2\gamma_3 t_0} & 0 & 0 & 1
\end{array}\right).
\end{equation}
It follows that ${\cal N}[{\cal Z}(\chi_0, t_0)] = e^{-2\gamma_3t_0}$.  Under the influence of a phase noise, the entanglement of $\chi_0$ decays 
smoothly and asymptotically exponentially to zero.  It does not reach zero in a finite time.  In an interesting paper \cite{Huang}, Huang and Zhu 
showed that to observe ESD under dephasing, an entangled two-qubit density operator should have nonzero diagonal elements.  We show below how this 
insight throws light on ESD discussed in Refs.\cite{Yeo, Yu}.

\subsection{Bit-flip and phase noise}
Now, if one of the particles is instead subject to a bit-flip noise described by $L_{01} = \sigma_0 \otimes \sigma_1$ (or $L_{10} = \sigma_1 \otimes 
\sigma_0$), then solving
$$
\frac{d}{dt}\rho = \gamma_1[(\sigma_0 \otimes \sigma_1)\rho(\sigma_0 \otimes \sigma_1) - \rho].
$$
we have
\begin{equation}\label{diagnot0}
{\cal X}(\chi_0, t_0) = 
\frac{1}{4}\left(\begin{array}{cccc}
1 + e^{-2\gamma_1 t_0} & 0 & 0 & 1 + e^{-2\gamma_1 t_0} \\
0 & 1 - e^{-2\gamma_1 t_0} & 1 - e^{-2\gamma_1 t_0} & 0 \\
0 & 1 - e^{-2\gamma_1 t_0} & 1 - e^{-2\gamma_1 t_0} & 0 \\
1 + e^{-2\gamma_1 t_0} & 0 & 0 & 1 + e^{-2\gamma_1 t_0}
\end{array}\right),
\end{equation}
which has nonzero diagonal elements.  Clearly, ${\cal N}[{\cal X}(\chi_0, t_0)] = e^{-2\gamma_1t_0}$.  So, under the influence of a bit-flip noise, the 
entanglement of $\chi_0$ decays smoothly and asymptotically exponentially to zero, in exactly the same manner as when it is under the influence of a 
phase noise.  This can be understood from the facts that $|0\rangle$ and $|1\rangle$ are eigenvectors of $\sigma_3$, and that $|\Psi^0_{\rm Bell}\rangle$ 
can also be similarly expressed in terms of the eigenvectors $|\pm\rangle \equiv (|0\rangle \pm |1\rangle)/\sqrt{2}$ of $\sigma_1$:
$$
|\Psi^0_{\rm Bell}\rangle = \frac{1}{\sqrt{2}}(|+\rangle|+\rangle + |-\rangle|-\rangle).
$$
Therefore, to the state $\chi_0$, a bit-flip noise is like a phase noise.

Such freedom is lost in the presence of both bit-flip and phase noise.  Consequently, we find a dramatically different effect when we subject $\chi_0$ 
simultaneously to $L_{01}$ and $L_{03}$.  The resulting state
\begin{equation}\label{xz}
{\cal XZ}(\chi_0, t_0) = \frac{1}{4}\left(\begin{array}{cccc}
1 + e^{-2\gamma_1 t_0} & 0 & 0 & e^{-2\gamma_3 t_0}(1 + e^{-2\gamma_1 t_0}) \\
0 & 1 - e^{-2\gamma_1 t_0} & e^{-2\gamma_3 t_0}(1 - e^{-2\gamma_1 t_0}) & 0 \\
0 & e^{-2\gamma_3 t_0}(1 - e^{-2\gamma_1 t_0}) & 1 - e^{-2\gamma_1 t_0} & 0 \\
e^{-2\gamma_3 t_0}(1 + e^{-2\gamma_1 t_0}) & 0 & 0 & 1 + e^{-2\gamma_1 t_0}
\end{array}\right)
\end{equation}
can be obtained by solving
$$
\frac{d}{dt}\rho = \gamma_1[(\sigma_0 \otimes \sigma_1)\rho(\sigma_0 \otimes \sigma_1) - \rho] + 
                   \gamma_3[(\sigma_0 \otimes \sigma_3)\rho(\sigma_0 \otimes \sigma_3) - \rho].
$$
Assuming for simplicity, $\gamma_k = \gamma$ from here on, ${\cal XZ}(\chi_0, t_0)$ has negativity \cite{remark1}
\begin{equation}\label{mu}
{\cal N}[{\cal XZ}(\chi_0, t_0)] = \max\left\{0, \frac{1}{2}[(1 + e^{-2\gamma t_0})^2 - 2]\right\}.
\end{equation}
The entanglement suffers sudden death in a finite time $\tau_{\rm d} = -\ln(\sqrt{2} - 1)/(2\gamma)$.

We highlight that lying at the root of ESD here is the noncommutativity of $\sigma_1$ and $\sigma_3$:
$$
[\sigma_3,\ \sigma_1] = 2i\sigma_2.
$$
Consequently, the noise generated by $\sigma_1$ causes zero diagonal elements like those in Eq.(\ref{diag0}) to become nonzero and hence ESD 
\cite{Huang}.  This is supported by the fact that $L_{02} = \sigma_0 \otimes \sigma_2$ with $L_{03}$ gives exactly the same result, Eq.(\ref{mu}).  We 
call noises that are generated by non-commuting operators non-commuting.  Lastly, we note that the depolarizing noise \cite{Nielsen}, which is generated 
by $\sigma_1$, $\sigma_2$, and $\sigma_3$, may thus cause ESD.  In fact, subjecting one of the particles in $\chi_0$ to a depolarizing noise yields a 
state with negativity given by $\max\{0, (3e^{-4\gamma t_0} - 1)/2\}$.  ESD occurs at $\tau_{\rm d} = \ln 3/(4\gamma)$.  Non-commuting Pauli noises 
therefore cause ESD in $\chi_0$.

Equations (\ref{diag0}), (\ref{diagnot0}) and (\ref{xz}) could describe the physical situation where Alice prepares, at time $t = 0$, the state $\chi_0$ 
and sends one of the particles down a noisy channel ${\cal T}$ to Bob to establish at $t_0$ the shared entangled state $\chi$.  We call $t_0$ the 
transmission time and say the mixed state $\chi = {\cal T}(\chi_0, t_0)$ results from transmission noise.

\subsection{Amplitude and phase noise}
In Ref.\cite{Yu}, the combined effect of amplitude and phase noise on a generic class of entangled states was studied.  Here, we wish to point out how 
the commutation relations for generators of amplitude and Pauli noise could account for ESD in $\chi_0$.  Firstly, we recall that the amplitude noise is 
described by the Lindblad operator
\begin{equation}
\sigma_- \equiv \frac{1}{2}(\sigma_1 + i\sigma_2).
\end{equation}
Solving
$$
\frac{d}{dt}\rho = \gamma[(\sigma_0 \otimes \sigma_-)\rho(\sigma_0 \otimes \sigma_-)^{\dagger} 
                 - \frac{1}{2}(\sigma_0 \otimes \sigma_-)^{\dagger}(\sigma_0 \otimes \sigma_-)\rho
                 - \frac{1}{2}\rho(\sigma_0 \otimes \sigma_-)^{\dagger}(\sigma_0 \otimes \sigma_-)],
$$
we obtain
\begin{equation}\label{Bob}
{\cal B}(\chi_0, t_0) = \frac{1}{2}\left(\begin{array}{cccc}
1 & 0 & 0 & e^{-\gamma t_0/2} \\
0 & 0 & 0 & 0 \\
0 & 0 & 1 - e^{-\gamma t_0} & 0 \\
e^{-\gamma t_0/2} & 0 & 0 & e^{-\gamma t_0}
\end{array}\right)
\end{equation}
with ${\cal N}[{\cal B}(\chi_0, t_0)] = e^{-\gamma t_0}$.  We note that, in contrast to ${\cal X}$ and ${\cal Z}$, $\sigma_- \otimes \sigma_0$ acting on 
$\chi_0$ produces a different state
\begin{equation}\label{Alice}
{\cal A}(\chi_0, t_0) = \frac{1}{2}\left(\begin{array}{cccc}
1 & 0 & 0 & e^{-\gamma t_0/2} \\
0 & 1 - e^{-\gamma t_0} & 0 & 0 \\
0 & 0 & 0 & 0 \\
e^{-\gamma t_0/2} & 0 & 0 & e^{-\gamma t_0}
\end{array}\right)
\end{equation}
with ${\cal N}[{\cal A}(\chi_0, t_0)] = e^{-\gamma t_0}$.  This asymmetry will lead to some complication that we will encounter later on.

Secondly, recall that $\sigma_-$ satisfies the commutation relations $[\sigma_-,\ \sigma_1] = \sigma_3$, $[\sigma_-,\ \sigma_2] = i\sigma_3$, and 
$[\sigma_3,\ \sigma_-] = 2\sigma_-$.  It follows from the above analysis that if we subject $\chi_0$ simultaneously to both $L_{01}$ 
and $L_{0a}$, where $L_{0a} \equiv \sigma_0 \otimes \sigma_-$, then we should observe ESD.  This is indeed the case.  Solving
\begin{eqnarray}
\frac{d}{dt}\rho & = & \gamma[(\sigma_0 \otimes \sigma_1)\rho(\sigma_0 \otimes \sigma_1) - \rho] + \nonumber \\
& &                    \gamma[(\sigma_0 \otimes \sigma_-)\rho(\sigma_0 \otimes \sigma_-)^{\dagger} 
                     - \frac{1}{2}(\sigma_0 \otimes \sigma_-)^{\dagger}(\sigma_0 \otimes \sigma_-)\rho
                     - \frac{1}{2}\rho(\sigma_0 \otimes \sigma_-)^{\dagger}(\sigma_0 \otimes \sigma_-)], \nonumber
\end{eqnarray}
we obtain
\begin{equation}
{\cal BX}(\chi_0, t_0) = \frac{1}{6}\left(\begin{array}{cccc}
2 + e^{-3\gamma t_0} & 0 & 0 & 3e^{-3\gamma t_0/2}\cosh\gamma t_0    \\
0 & 1 - e^{-3\gamma t_0} & 3e^{-3\gamma t_0/2}\sinh\gamma t_0 & 0  \\
0 & 3e^{-3\gamma t_0/2}\sinh\gamma t_0 & 2(1 - e^{-3\gamma t_0}) & 0 \\
3e^{-3\gamma t_0/2}\cosh\gamma t_0 & 0 & 0 & 1 + 2e^{-3\gamma t_0}
\end{array}\right),
\end{equation}
which has negativity
\begin{equation}\label{nu}
{\cal N}[{\cal BX}(\chi_0, t_0)] = \max\left\{0, \frac{1}{6}\left[3e^{-3\gamma t_0} + 
\sqrt{2}e^{-3\gamma t_0/2}\sqrt{8 + 9\cosh(2\gamma t_0) + \cosh(3\gamma t_0)} - 3\right]\right\}.
\end{equation}
The entanglement becomes zero in a finite time $\tau_{\rm d} \approx 0.747282/\gamma$, which is a solution of
$$
8\cosh3\gamma t - 9\cosh2\gamma t - 17 = 0.
$$
$L_{02}$ with $L_{0a}$ gives exactly the same result, Eq.(\ref{nu}).  But, for $\chi_0$, $L_{03}$ and $L_{0a}$ do not result in ESD despite the fact 
that they do not commute.  They produce a state with negativity given by $[\sqrt{(1 - e^{-\gamma t})^2 + 4e^{-5\gamma t}} - (1 - e^{-\gamma t})]/2$.  
This is consistent with the result in \cite{Huang}, since $L_{03}$ and $L_{0a}$ acting on $\chi_0$ do not produce a state with nonzero diagonal elements.  
So, non-commutativity between the Lindblad generators is a necessary but not sufficient condition for $\chi_0$ to suffer ESD.  However, in realistic 
superdense coding, we will encounter the following state
\begin{equation}
{\cal AZ}({\cal BZ}(\chi_0, t_0), t_0) = 
\frac{1}{2}\left(\begin{array}{cccc}
1 + (1 - e^{-\gamma t_0})^2 & 0 & 0 & e^{-5\gamma t_0} \\
0 & e^{-\gamma t_0}(1 - e^{-\gamma t_0}) & 0 & 0 \\
0 & 0 & e^{-\gamma t_0}(1 - e^{-\gamma t_0}) & 0 \\
e^{-5\gamma t_0} & 0 & 0 & e^{-2\gamma t_0}\end{array}\right),
\end{equation}
which suffers ESD.  Its negativity is given by $\max\{0, e^{-\gamma t_0}(e^{-\gamma t_0} + e^{-4\gamma t_0} - 1)\}$ and becomes zero at $\tau_{\rm d}$ 
that can be obtained from a solution of
$$
e^{-\gamma t} + e^{-4\gamma t} - 1 = 0.
$$
We have $\tau_d \approx 0.644569/\gamma$.

\subsection{Single-qubit rotations in the presence of phase noise}
Single-qubit rotations are essential to quantum information processing.  Here, we consider Hamiltonians  given by $H_m \equiv \omega_0\sigma_m/2$, which 
generates an anticlockwise coherent rotation of a qubit about the $m$-axis at the rate $\omega_0$.  $H_m$ describes rather generically the dynamics of 
two-level systems.  For example, it could describe an electron spin in an external magnetic field, or it could describe a two-level atom in a light 
field.  From the above insights we expect, for instance, $\chi_0$ to suffer ESD when it is under the action of $H_{10}$ or $H_{20}$ in the presence of 
noise generated by $L_{30}$.  Here, $H_{m0} = H_m \otimes \sigma_0$.  To show that this is indeed the case, we solve
$$
\frac{d}{dt}\rho = -i[H_{m0}, \rho] + \gamma[(\sigma_3 \otimes \sigma_0)\rho(\sigma_3 \otimes \sigma_0) - \rho]
$$
$L_{30}$ together with different $H_{m0}$'s, produce different noisy single-qubit rotations ${\cal R}^{(p)}_m$.  We find the entanglement of 
${\cal R}^{(p)}_1(\chi_0, t)$ goes to zero in a finite time $\tau_d$, which is a solution of
\begin{equation}
e^{-2\gamma t} + \frac{2e^{-\gamma t}}{\omega}\sqrt{\omega^2_0 - \gamma^2\cos^2\omega t} - 1 = 0.
\end{equation}
Here, $\omega \equiv \sqrt{\omega^2_0 - \gamma^2}$.  We observe that for a fixed $\gamma$, interestingly $\tau_d$ decreases with increasing $\omega_0$.  
And, in the limit of large $\omega_0$, $\tau_{\rm d} \approx -\ln(\sqrt{2} - 1)/\gamma$.  This is intriguing since it is intuitive to reduce the impact 
of noise by increasing $\omega_0$ so as to reduce the time taken to complete a desired rotation.
\begin{table}
\begin{tabular}{|c|c|c|c|c|c|c|}
\hline
$\omega_0$                   & $10^{-2}$ & $10^{-1}$ & $10^0$  & $10^1$  & $10^2$  & $10^3$  \\ \hline
$\tau_{\rm d}(\gamma = 0.1)$ & 27.2453   & 12.4897   & 8.82654 & 8.81375 & 8.81374 & 8.81374 \\ \hline
$\tau_{\rm d}(\gamma = 0.2)$ & 16.3894   & 8.05990   & 4.47185 & 4.40687 & 4.40687 & 4.40687 \\ \hline
\end{tabular}
\caption{$\tau_{\rm d}$'s of ${\cal R}^{(p)}_1(\chi_0, t)$ for different $\omega_0$'s, when $\gamma = 0.1$ or $\gamma = 0.2$.  For a fixed $\omega_0$, 
$\tau_{\rm d}$ increases with increasing $\gamma$.}
\end{table}
TABLE I shows the dependence of $\tau_{\rm d}$, for ${\cal R}^{(p)}_1(\chi_0, t)$, on $\omega_0$.  We obtain exactly the same conclusions with 
${\cal R}^{(p)}_2(\chi_0, t)$, but not with ${\cal R}^{(p)}_3(\chi_0, t)$.  The negativity of ${\cal R}^{(p)}_3(\chi_0, t)$ is given by 
$e^{-2\gamma t}$, independent of $\omega_0$.  This is because in this case the Lindblad generator of phase noise commutes with the Hamiltonian.  
Non-commutativity between generators of noises or unitary rotation and noise is necessary for ESD to occur.  This is further illustrated in the 
following examples.

\subsection{Single-qubit rotations in the presence of amplitude noise}
Consider ${\cal R}^{(a)}_1(\chi_0, t)$, which is obtained by solving
\begin{eqnarray}
\frac{d}{dt}\rho & = & -i[H_{m0}, \rho] + \gamma[(\sigma_- \otimes \sigma_0)\rho(\sigma_- \otimes \sigma_0)^{\dagger} - \nonumber \\
& &                    \frac{1}{2}(\sigma_- \otimes \sigma_0)^{\dagger}(\sigma_- \otimes \sigma_0)\rho
                     - \frac{1}{2}\rho(\sigma_- \otimes \sigma_0)^{\dagger}(\sigma_- \otimes \sigma_0)], \nonumber
\end{eqnarray}
with $\chi_0$ as the initial state.  In contrast to the above subsection, we are not able to obtain a closed expression for the negativity of 
${\cal R}^{(a)}_1(\chi_0, t)$.  Numerical results show that $\tau_{\rm d}$'s for ${\cal R}^{(a)}_1(\chi_0, t)$ behave similarly to those for 
${\cal R}^{(p)}_1(\chi_0, t)$.  Namely, we observe that for a fixed $\gamma$, $\tau_{\rm d}$ decreases with increasing $\omega_0$.  And, in the limit of 
large $\omega_0$, it becomes a constant that decreases with increasing $\gamma$.  TABLE II shows the dependence of $\tau_{\rm d}$, for 
${\cal R}^{(a)}_1(\chi_0, t)$, on $\omega_0$.
\begin{table}
\begin{tabular}{|c|c|c|c|c|c|c|}
\hline
$\omega_0$                   & $10^{-2}$ & $10^{-1}$ & $10^0$  & $10^1$  & $10^2$  & $10^3$  \\ \hline
$\tau_{\rm d}(\gamma = 0.1)$ & 93.2803   & 29.1270   & 16.9282 & 16.7860 & 16.7847 & 16.7847 \\ \hline
$\tau_{\rm d}(\gamma = 0.2)$ & 60.0374   & 21.3006   & 8.70814 & 8.39552 & 8.39239 & 8.35235 \\ \hline
\end{tabular}
\caption{$\tau_{\rm d}$'s of ${\cal R}^{(a)}_1(\chi_0, t)$ for different $\omega_0$'s, when $\gamma = 0.1$ and $\gamma = 0.2$.}
\end{table}
We remark that while ${\cal R}^{(a)}_2(\chi_0, t)$ yields exactly the same results, the negativity of ${\cal R}^{(a)}_3(\chi_0, t)$ is given by 
$e^{-\gamma t}$, again independent of $\omega_0$.  This is consistent with the conclusions at the end of subsection C.

In summary, noisy quantum operations that result from non-commuting Hamiltonian and Lindblad operator could have an equally devastating effect on the 
entanglement of $\chi_0$ as non-commuting noises.  The consequence of these results in ${\cal S}$ will be discussed after we briefly review realistic 
special dense coding in the following section.

\section{Superdense coding}
\subsection{Holevo function}
Suppose Alice communicates some classical information with Bob using ``alphabet'' states $x_1, x_2, x_3, \cdots$ with probabilities $p_1, p_2, p_3, 
\cdots$ respectively.  The average number of bits of classical information Alice can convey to Bob per transmission of an alphabet state is bounded from 
above by the Holevo function \cite{Holevo}
\begin{equation}\label{holevo}
H[x] = S[x] - \sum_ip_iS[x_i],
\end{equation}
where $x = \sum_ip_ix_i$.  This bound can be achieved in the limit of an infinite ensemble by appropriate block coding at Alice's end and appropriate 
measurements at Bob's end \cite{Schumacher}.

\subsection{Realistic special dense coding}
Throughout Section IV, we suppose Alice prepares pairs of qubits in the state $\chi_0$ at time $t = 0$.  She then sends one of each pair of qubits via a 
noisy channel ${\cal T}$, which reaches Bob in time $t_0$.  In realistic special dense coding, Alice and Bob therefore initially share pairs of 
entangled qubits in some mixed state $\chi = {\cal T}(\chi_0, t_0)$.  During this process, we assume Alice's other qubit is well shielded from 
environmental influences.  After $\chi$ has been established, Alice generates the following four letter states with equal probability:
\begin{eqnarray}
\alpha_0 & = & (\sigma_0 \otimes \sigma_0)\chi(\sigma_0 \otimes \sigma_0), \nonumber \\
\alpha_m & = & {\cal R}^{(\eta)}_m(\chi),                                  \nonumber
\end{eqnarray}
where $\eta$ indicates the type of environmental noise affecting the transmissions and encoding operations \cite{remark}.  We assume Bob's qubits are 
well isolated throughout Alice's noisy encoding.  For $\alpha_0$, the qubits are perfectly isolated since no operations need to be performed on Alice's 
qubit.  The environment will only affect Alice's qubit during a nontrivial quantum operation.  Lastly, Alice sends her encoded qubits to Bob, which for 
simplicity we suppose is done via the same noisy channel ${\cal T}$ in time $t_0$.  By so doing, Alice is essentially commnicating with Bob using the 
alphabet states
\begin{eqnarray}
x_0 & = & {\cal T}(\alpha_0, t_0), \nonumber \\
x_m & = & {\cal T}(\alpha_m, t_0)
\end{eqnarray}
as separate letters.  The number of bits she can communicate with Bob using this procedure is thus bounded by the Holevo function,
\begin{equation}\label{sdcHol}
H^{(\eta)}[x] = S[x] - \frac{1}{4}\sum^3_{m = 0}S[x_m],
\end{equation}
where $x = 1/4\sum^3_{m = 0}x_m$.  We stress that only when a qubit is being manipulated (rotated or transmitted), then it will be affected by the 
environment.  Otherwise, the qubits are assumed to be perfectly isolated.  In addition, we have assumed that Bob's joint measurements necessary to 
decode Alice's message are perfect and do not result in further loss in information to the environment.  Ideally, Alice and Bob will initially share 
pairs of qubits in the maximally entangled state $\chi_0$, $x_m = (\sigma_m \otimes \sigma_0)\chi_0(\sigma_m \otimes \sigma_0)$, and $S[x_m] = 0$.  
Also, $x = 1/4I$ with $I$ the four-dimensional identity.  In this case, it follows that $H[x] = 2$.  This is the ideal classical information capacity of 
superdense coding \cite{Bennett92}.  In the following section, we present our analysis of the impact of non-commuting noises and noisy encoding 
operations on $\cal S$.

\section{Environmental effects on special dense coding}
For clarity we consider noisy transmission and noisy encoding 
separately, before discussing the cases where both transmission and encoding are noisy.  In subsections A and B, Alice's encoding operations are ideal 
and unitary, but the qubits are sent via noisy Pauli channels.  We find that when environmental influences are capable of causing ESD in $\chi_0$, they 
may result in $\chi_0$ losing its capacity to do dense coding before ESD.  From the results in Section II we know that noisy Pauli rotations could 
also lead to ESD in $\chi_0$.  These operations are basic to ${\cal S}$.  We show, in subsection C, that Alice's noisy encoding operations alone could 
indeed result in an equally devastating impact on ${\cal S}$.  These results are in contrast to those for standard teleportation \cite{Yeo}, and the 
implications will be briefly discussed in subsection B.

\subsection{Phase noise}
Alice establishes the shared entangled state $\chi$ with Bob via a dephasing channel, i.e., ${\cal T} = {\cal Z}$ and $\chi = {\cal Z}(\chi_0, t_0)$.  
It follows that
\begin{eqnarray}
x_0 & = & {\cal Z}((\sigma_0 \otimes \sigma_0)\chi(\sigma_0 \otimes \sigma_0), t_0), \nonumber \\
x_m & = & {\cal Z}((\sigma_m \otimes \sigma_0)\chi(\sigma_m \otimes \sigma_0), t_0).
\end{eqnarray}
Note that since $L_{03}$ and $L_{30}$ give the same ${\cal Z}(\chi_0, t_0)$, namely Eq.(\ref{diag0}), we have $x_0 = {\cal Z}(\chi_0, 2t_0)$ and 
$x_m = (\sigma_m \otimes \sigma_0){\cal Z}(\chi_0, 2t_0)(\sigma_m \otimes \sigma_0)$.  This is mathematically equivalent to the situation considered by 
Bose, {\em et al.} \cite{Bose} and Bowen \cite{Bowen}.  Consequently, the classical information capacity of dense coding is given by Eq.(\ref{cap}),
\begin{equation}
{\cal C}[{\cal Z}(\chi_0, 2t_0)] = 2 + \frac{1}{2}(1 - e^{-4\gamma t_0})\log_2[\frac{1}{2}(1 - e^{-4\gamma t_0})] 
                                     + \frac{1}{2}(1 + e^{-4\gamma t_0})\log_2[\frac{1}{2}(1 + e^{-4\gamma t_0})].
\end{equation}
${\cal C}[{\cal Z}(\chi_0, 2t_0)]$ decays smoothly and asymptotically exponentially to the limiting ``classical'' value of $1$.  So, provided Alices's 
encoding operations are perfect, every bit of entanglement associated with ${\cal Z}(\chi_0, 2t_0)$ will yield ``nonclassical'' classical information 
capacity, albeit one that decreases with increasing $t_0$.  Lastly, we note that ${\cal C}[{\cal X}(\chi_0, 2t_0)] = {\cal C}[{\cal Z}(\chi_0, 2t_0)]$.

\subsection{Bit flip and phase noise}
If the noisy quantum channel is a combination of bit-flip and phase noise, i.e., ${\cal T} = {\cal XZ}$ and $\chi = {\cal XZ}(\chi_0, t_0)$, then
\begin{eqnarray}
x_0 & = & {\cal XZ}((\sigma_0 \otimes \sigma_0)\chi(\sigma_0 \otimes \sigma_0), t_0), \nonumber \\
x_m & = & {\cal XZ}((\sigma_m \otimes \sigma_0)\chi(\sigma_m \otimes \sigma_0), t_0).
\end{eqnarray}
Again, since $L_{01}$, $L_{03}$ and $L_{10}$, $L_{30}$ give the same ${\cal XZ}(\chi_0, t_0)$, namely Eq.{\ref{xz}}, we have 
$x_0 = {\cal XZ}(\chi_0, 2t_0)$ and $x_m = (\sigma_m \otimes \sigma_0){\cal XZ}(\chi_0, 2t_0)(\sigma_m \otimes \sigma_0)$, and
\begin{eqnarray}\label{capdeath}
{\cal C}[{\cal XZ}(\chi_0, 2t_0)] & = & 2 + \frac{1}{2}(1 - e^{-8\gamma t_0})\log_2[\frac{1}{4}(1 - e^{-8\gamma t_0})]   \nonumber \\
& &                                       + \frac{1}{2}(1 - e^{-4\gamma t_0})^2\log_2[\frac{1}{2}(1 - e^{-4\gamma t_0})] \nonumber \\
& &                                       + \frac{1}{2}(1 + e^{-4\gamma t_0})^2\log_2[\frac{1}{2}(1 + e^{-4\gamma t_0})].
\end{eqnarray}
In contrast to ${\cal X}(\chi_0, 2t_0)$ and ${\cal Z}(\chi_0, 2t_0)$, there exists a critical total transmission time $\tau_{\rm c}$ beyond which we 
will fail to have superdense coding.  $\tau_{\rm c}$ depends on $\gamma$ and is given by $\tau_{\rm c} \approx 0.124266/\gamma$.  This is roughly a 
factor of four times smaller than $\tau_d$.  The state ${\cal XZ}(\chi_0, \tau_{\rm c})$ thus fails to be useful for superdense dense coding way before 
its entanglement becomes zero.  The nonzero entanglement of ${\cal XZ}(\chi_0, \tau_{\rm c})$ does not enable $\cal S$ to yield ``nonclassical" classical 
information capacity. By analogy with ESD, we could call this sudden death of special dense coding.  This is clearly different from ESD as the 
entanglement associated with the state ${\cal XZ}(\chi_0, \tau_{\rm c})$ is not totally destroyed.

Intriguingly, ${\cal XZ}(\chi_0, t)$ only fails to yield non-classical teleportation fidelity at exactly the moment $\tau_{\rm d}$ when its entanglement 
suffers a sudden death \cite{Yeo}.  That is, every single bit of entanglement associated with ${\cal XZ}(\chi_0, t)$ is useful for teleportation.  
Although there is no doubt entanglement is a necessary resource for both superdense coding and teleportation, this example shows that different quantum 
protocols could depend more or less critically on some possibly different features of multipartite quantum systems.  These could perhaps be aspects of 
multipartite quantum systems not explored before.  We shall not pursue this issue further here.  Instead, we note that if the noisy quantum channel is 
a combination of amplitude and phase noise, then $\chi = {\cal BZ}(\chi_0, t_0)$ and
\begin{eqnarray}\label{amphase}
x_0 & = & {\cal AZ}((\sigma_0 \otimes \sigma_0)\chi(\sigma_0 \otimes \sigma_0), t_0), \nonumber \\
x_m & = & {\cal AZ}((\sigma_m \otimes \sigma_0)\chi(\sigma_m \otimes \sigma_0), t_0).
\end{eqnarray}
For this case, we determine $H^{(a)}[x]$, Eq.(\ref{sdcHol}).  This is due to the asymmetry, which we pointed out earlier in Eqs.(\ref{Bob}) and 
(\ref{Alice}).  Consequently, this case does not reduce formally to that considered by Bose, {\em et al}. \cite{Bose} and Bowen \cite{Bowen}.  To claim 
that what we have calculated is the classical information capacity may require further work beyond the scope of this paper.  However, we wish to 
highlight that in the limit $\gamma \rightarrow 0$ or $t_0 \rightarrow 0$ we do have $H^{(a)}[x] \rightarrow 2$.  Furthermore, the critical total 
transmission time $\tau_{\rm c}$ beyond which $H^{(a)}[x]$ falls below $1$ is given by $\tau_{\rm c} \approx 0.172879/\gamma$, which is 
again roughly a factor of four times smaller than the corresponding $\tau_d$.  Our numerical studies thus give similar results and hence conclusions as 
above.  We believe that the non-commutativity between generators of the transmission noises is responsible for both ESD and sudden death of $\cal S$ 
here.  This is further supported by the examples below.

\begin{figure}[b]
{\includegraphics{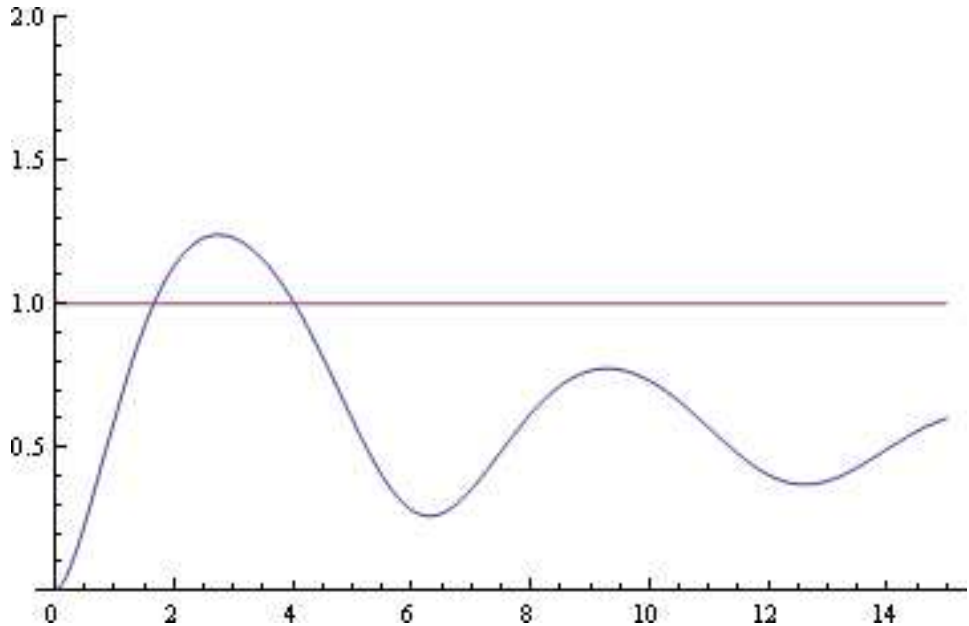}}
\caption{$H^{(p)}[x]$ versus encoding time $t$ for $\gamma = 1/10$ and $\omega_0 = 1$.}
\end{figure}

\begin{figure}[t]
{\includegraphics{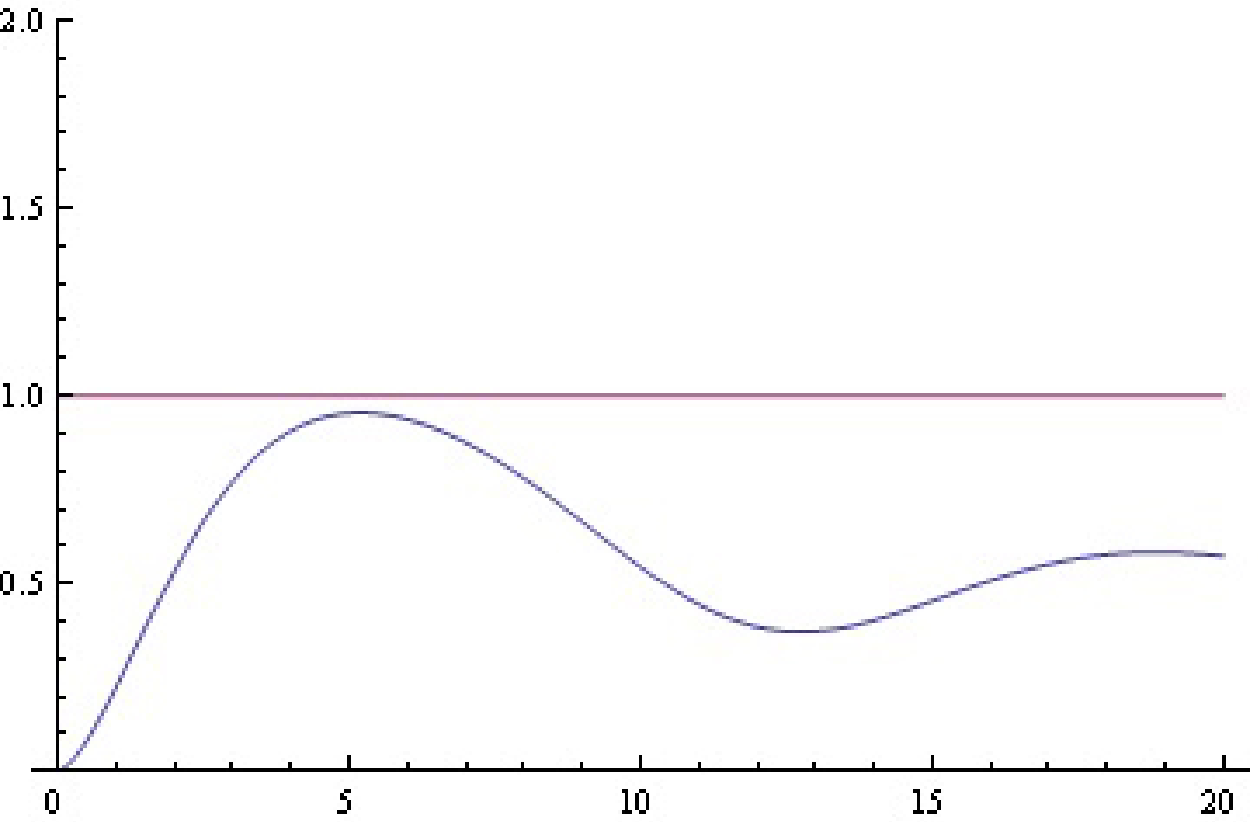}}
\caption{$H^{(p)}[x]$ versus encoding time $t$ for $\gamma = 1/10$ and $\omega_0 = 1/2$.}
\end{figure}

\begin{figure}[t]
{\includegraphics{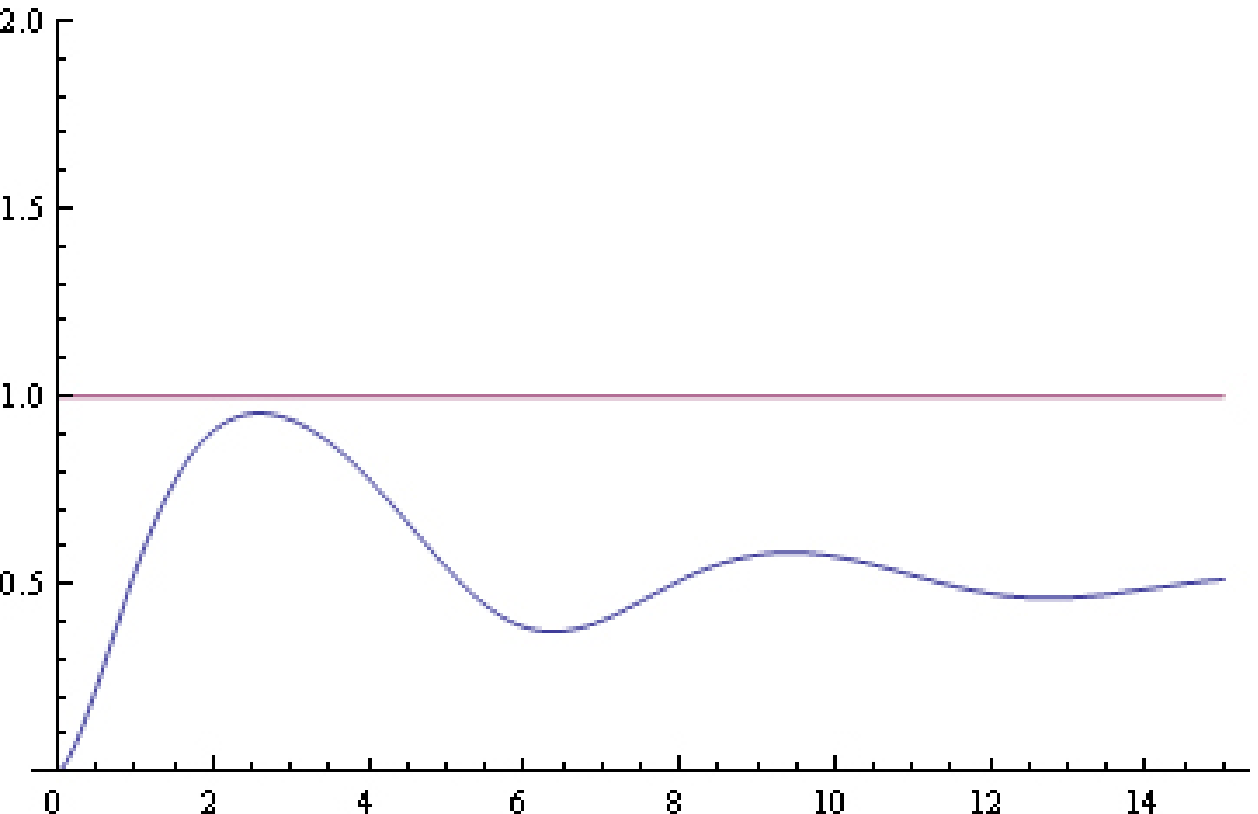}}
\caption{$H^{(p)}[x]$ versus encoding time $t$ for $\gamma = 1/5$ and $\omega_0 = 1$.}
\end{figure}

\begin{figure}[b]
{\includegraphics{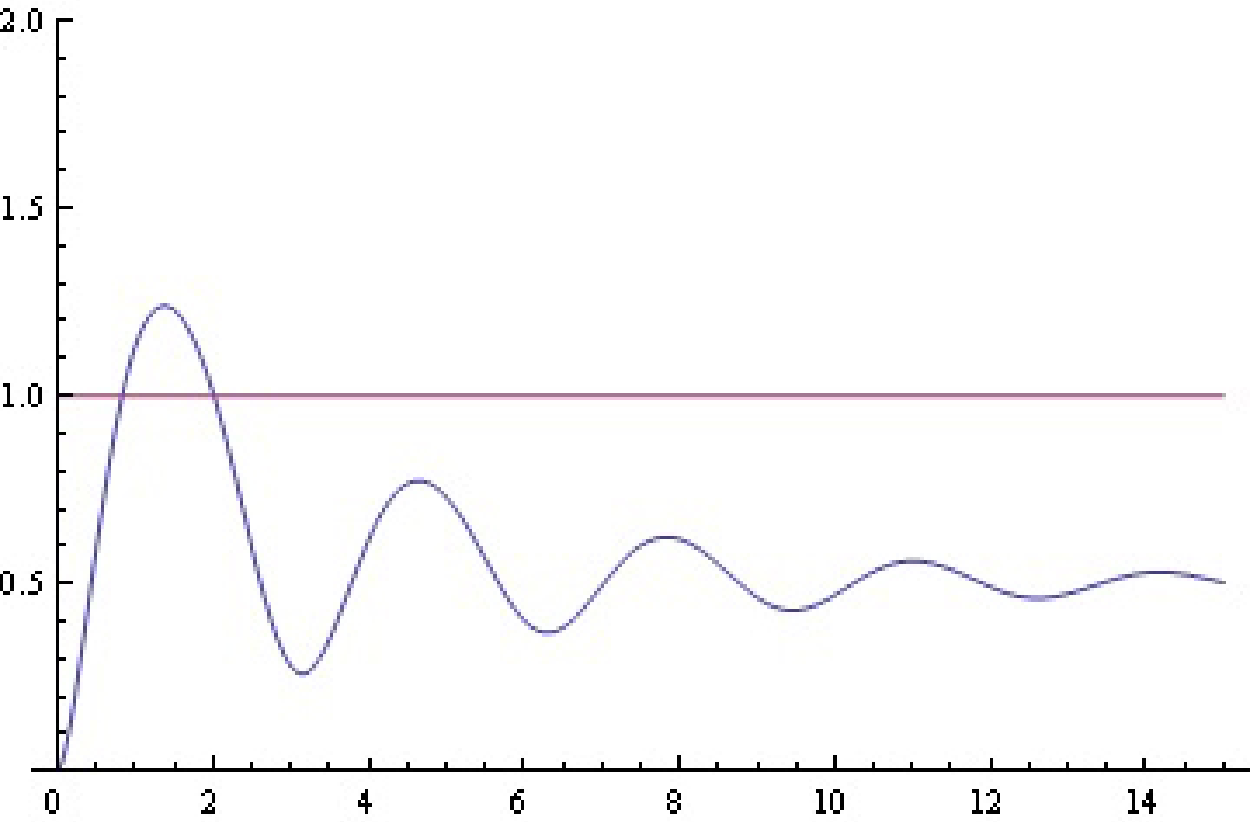}}
\caption{$H^{(p)}[x]$ versus encoding time $t$ for $\gamma = 1/5$ and $\omega_0 = 2$.}
\end{figure}

\subsection{Single-qubit rotations in the presence of noise}
Let us now suppose the transmission is noiseless and Alice shares $\chi_0$ with Bob, while Alice's encoding operations are corrupted by phase noise.  
We thus have
\begin{eqnarray}
x_0 & = & \chi_0, \nonumber \\
x_m & = & {\cal R}^{(p)}_m(\chi_0).
\end{eqnarray}
Like Eq.(\ref{amphase}), this is clearly different from the situation considered by Bose, {\em et al.} and Bowen.  However, since we are interested in 
the impact of Alice's noisy encoding on $\cal S$, we calculate
\begin{equation}
H^{(p)}[x] = S[x] - \frac{1}{4}\sum^3_{m = 1}S[x_m].
\end{equation}
Again, we observe that $H^{(p)}[x] \rightarrow 2$ as $\gamma \rightarrow 0$, independent of $\omega_0$.  It is straightforward to show that 
$H^{(b)}[x] = H^{(bp)}[x] = H^{(p)}[x]$, which is in general a very complicated function of $\gamma$, $\omega_0$, and encoding time $t$.  Figure 1 
shows $H^{(p)}[x]$ as a function of encoding time $t$ when $\gamma = 1/10$ and $\omega_0 = 1$.  The first maximum $H^{(p)}_{\max}[x]$ is 
approximately $1.23928$ and is achieved at time $T^{(p)}_{\chi_0}$ roughly equal to $2.75085$.  Subsequent maxima fall below 1, the ``classical'' limit.  
Figure 2 shows the case when $\gamma = 1/10$ but $\omega_0 = 1/2$.  Now, even the first maximum falls below 1.  Indeed, there exists a critical 
$\omega_0$, denoted by $\omega^{(p)}_c[\chi_0]$, that is approximately $0.56$ below which the first maximum will fall below 1.  This maximum is attained 
at $\tau^{(p)}_c[\chi_0] \approx 4.68027$.  A smaller $\omega_0$ means a longer duration for the encoding to be completed.  And, in the presence of 
noise, this means a further degradation of the information transfer.  For a larger $\gamma$, we will have to demand a higher critical 
$\omega^{(p)}_c[\chi_0]$.  This is demonstrated in Figures 3 and 4 where $\gamma = 1/5$, and $\omega_0 = 1$ and $2$ respectively.

We emphasize what is intriguing is that at the critical time $\tau^{(p)}_c[\chi_0]$ the entanglement of the state shared between Alice and Bob is 
nonzero (see TABLE I).  Therefore, in the presence of noise there is a constraint on the rate at which Alice's operations have to be executed in order 
for ${\cal S}$ to yield ``nonclassical'' classical information transfer.  The existence of a critical $\omega^{(p)}_c[\chi_0]$ means Alice's operations 
have to be completed in a finite time $\tau^{(p)}_c[\chi_0]$.  We have thus shown that even in the absence of transmission noise (i.e., noisy encoding 
alone) there is a finite lifetime beyond which ${\cal S}$ will fail to attain an information transfer better than classically possible.  Like 
Eq.(\ref{capdeath}), this is in contrast to the results for standard teleportation in Ref.\cite{Yeo}.  So, non-commuting Hamiltonians and Lindblad 
operators do have a bearing on the information transfer of $\cal S$.  If Alice's encoding operations are affected by amplitude noise instead, then we 
have $x_0 = \chi_0$ and $x_m = {\cal R}^{(a)}_m(\chi_0)$.  Similar numerical results are obtained and will not be presented here.

\subsection{Noisy encoding and transmission}
If in addition there is transmission noise ${\cal T}$, Alice and Bob will share the mixed state $\chi = {\cal T}(\chi_0, t_0)$, which has less 
entanglement to begin before Alice's noisy encoding operations.  The Holevo function $H^{(\eta)}[x]$ is a very complicated function of $\gamma$, 
$\omega_0$, $t_0$, and encoding time $t$.  We expect from the above results that for a fixed $\gamma$ and $t_0$ there will be a critical $\omega_0$ 
below which ${\cal S}$ will fail to achieve ``nonclassical'' classical information transfer.  Furthermore, depending on $\gamma$ and $t_0$, this 
critical $\omega_0$ must be correspondingly higher than that above.  Our numerical results indicate that this is indeed the case.  It is certainly more 
technically challenging in realistic situations.

\section{Conclusions}
A thorough understanding of all possible environmental effects, especially on quantum communication and computation, is obviously vital.  We would then 
be able to formulate effective strategies to overcome the challenges due to these detrimental effects.  In this paper, we have shown how besides 
non-commuting Lindblad operators, a Hamiltonian together with a non-commuting Lindblad operator produces a noisy quantum operation that could totally 
destroy the entanglement of a Bell state in a finite time.  More importantly, we have shown that when two quantum noises are capable of causing ESD in a 
Bell state, the resulting state becomes useless for $\cal S$ before ESD occurs.  We have also shown that when Alice's encoding operations are noisy a 
Bell state may fail to yield information transfer better than classically possible.  This can be accounted for by the fact that these operations could 
cause ESD in a Bell state.  These results are clearly in contrast to those in Ref.\cite{Yeo}.  We hope our finding will stimulate more complete and 
realistic studies on quantum information protocols under the influence of noise.


\begin{thebibliography}{99}
\bibitem{Nielsen} M. A. Nielsen and I. L. Chuang, {\em Quantum Computation and Quantum Information} (Cambridge University Press, Cambridge, 2000).
\bibitem{Bennett92} C. H. Bennett and S. J. Wiesner, Phys. Rev. Lett. {\bf 69}, 2881 (1992).

\bibitem{Breuer} H.-P. Breuer and F. Petruccione, {\it Theory of Open Quantum Systems} (Oxford University Press, Oxford, 2002).

\bibitem{Bose} S. Bose, M. Plenio, and V. Vedral, J. Mod. Opt. {\bf 47}, 291 (2000).
\bibitem{Bowen} G. Bowen, Phys. Rev. A {\bf 63}, 022302 (2001).

\bibitem{Yeo} Y. Yeo, Z.-W. Kho, and L. Wang, EPL {\rm 86}, 40009 (2009).

\bibitem{Yu} T. Yu and J. H. Eberly, Phys. Rev. Lett. {\bf 97}, 140403 (2006).
\bibitem{Almeida} M. P. Almedia, {\em et al.}, Science, {\bf 316}, 579 (2007).
\bibitem{Eberly} J. H. Eberly and T. Yu, Science {\bf 316}, 555 (2007).

\bibitem{Bennett93} C. H. Bennett, G. Brassard, C. Crepeau, R. Jozsa, A. Peres, and W. K. Wootters, Phys. Rev. Lett. {\bf 70}, 1895 (1993).

\bibitem{Peres} A. Peres, Phys. Rev. Lett. {\bf 77}, 1413 (1996).
\bibitem{Vidal} G. Vidal and R. F. Werner, Phys. Rev. A {\bf 65}, 032314 (2002).

\bibitem{Huang} J.-H. Huang and S.-Y. Zhu, Phys. Rev. A {\bf 76}, 062322 (2007).

\bibitem{remark1} The same result holds when $|\Psi^0_{\rm Bell}\rangle$ is subject to $L_{10} = \sigma^1 \otimes \sigma^0$ and $L_{03}$.  It is also 
not necessary to subject the state to the noises simultaneously.

\bibitem{Holevo} A. S. Kholevo, Probl. Peredachi Inf. {\bf 9}, 3 (1973) [Probl. Inf. Transm. {\bf 9}, 177 (1973)].
\bibitem{Schumacher} B. Schumacher and M. D. Westmoreland, Phys. Rev. A {\bf 56}, 131 (1997).

\bibitem{remark} For simplicity, we have supposed the same environmental noise during transmissions and encoding operations.
\end{thebibliography}
\end{document}